\documentclass[preprint]{revtex4-2}
\usepackage{amsmath,amssymb}
\usepackage{graphicx}
\usepackage{physics}
\usepackage{natbib}
\usepackage{amsfonts}
\usepackage{lineno}

\begin{document}
\title{Percolation of nonequilibrium assemblies of colloidal particles in active chiral liquids}

\author{Pragya Kushwaha$^{1}$, Sayan Maity$^1$, Anjaly S. Menon$^2$, Raghunath Chelakkot$^3$, Vijayakumar Chikkadi$^{1}$}

\affiliation{
$^{1}$ Indian Institute of Science Education and Research Pune, Pune 411008, India\\
$^{2}$ Indian Institute of Science Education and Research Tirupati, Tirupati 517619, India\\
$^{3}$ Indian Institute of Technology Bombay, Powai, Mumbai 400076, India\\
}

\begin{abstract}
The growing interest in the non-equilibrium assembly of colloidal particles in active liquids is driven by the motivation to create novel structures endowed with tunable properties unattainable within the confines of equilibrium systems. Here, we present an experimental investigation of the structural features of colloidal assemblies in active liquids of chiral E. coli. The colloidal particles form dynamic clusters due to the effective interaction mediated by active media. The activity and chirality of the swimmers strongly influence the dynamics and local ordering of colloidal particles, resulting in clusters with persistent rotation, whose structure differs significantly from those in equilibrium systems with attractive interactions, such as colloid-polymer mixtures. The colloid-bacteria mixture displays several hallmark features of a percolation transition at a critical density, where the clusters span the system size. However, a closer examination of the critical exponents associated with cluster size distribution, average cluster size, and correlation length in the vicinity of the critical density suggest strong deviations from the prediction of the standard continuum percolation model. Therefore, our experiments reveal a richer phase behavior of colloidal assemblies in active liquids.
\end{abstract}

\date{\today}

\maketitle
\section{Introduction}

A defining characteristic of active matter is the internal energy source that drives the system out of equilibrium\cite{Simha13, Volpe16}. Various living and synthetic systems, spanning a wide range of length scales and featuring diverse interaction rules among their elementary constituents, fall within this category. Examples include living systems, ranging from animals to bacteria, and synthetic systems such as microtubules driven by molecular motors \cite{Leibler97}, suspensions of Janus particles \cite{Bocquet10} or Quincke rollers \cite{Bartolo13}, and others. The distinguishing feature of active systems lies in their departure from the principles of detailed balance and time reversal symmetry. This deviation from equilibrium constraints gives rise to fascinating emergent phenomena, such as flocking \cite{Vicsek95, Bartolo13, Bartolo15}, motility-induced phase separation \cite{Cates15, Speck13, Palacci13}, active turbulence \cite{Goldstein12}, and superfluidity \cite{Clement15}. Nonequilibrium fluctuations in active systems have been exploited for various applications, including the self-assembly of novel biomimetic materials with diverse functionalities and drug delivery, among others \cite{Volpe16}.

In recent years, there has been a growing interest in understanding the phase behavior of mixtures involving both active and passive components, particularly emphasizing the phase behavior of passive entities in active liquids. Earlier experiments investigated the hierarchical organization of bio-polymers driven by molecular motors, leading to considerable insight into active gels \cite{Bausch11, Weitz09, Koenderink13, Dogic12}. These materials are inherently out of equilibrium, mimicking the nonlinear mechanical properties of cellular materials. Such studies have inspired the investigation of colloidal analogs of active gels. However, experiments on this topic are limited \cite{Palacci23}. The aspects that have received attention include the transport properties of colloidal particles at dilute densities \cite{Libchaber00, Poon13,  Clement11, Marenduzzo14} and the effective interaction between a pair of colloidal particles in active fluids \cite{Leonardo11, Naji17, Zhao21, Cacciuto14, Yang20, Peng23}. This has led to a better understanding of the interaction between passive and active particles. Despite these advances, the collective behavior of dense assemblies of colloidal particles in active liquids remains poorly explored in experimental systems. Recent experiments have reported dynamic clustering of colloidal particles in active liquids of E. coli suspensions \cite{Fakhri22, Chikkadi23, Auradou23}. The structure and size of the aggregates were reported to depend on the size ratio of colloidal particles to active particles \cite{Mishra18, Chikkadi23} and the density of active particles \cite{Fakhri22, Auradou23}. Earlier studies have revealed that E. coli swims in clockwise circular trajectories close to solid boundaries \cite{Stone06, Whitesides05}, and their chiral motion leads to interesting effects, such as the persistent rotational motion in microscopic gear-like structures suspended in such liquids \cite{Leonardo10, Aranson10}. A recent study had investigated the effect of chiral liquids on the phase behavior of sticky colloidal beads. The colloids were shown to form unconventional gel structures endowed with novel properties not achievable in equilibrium analogs \cite{Palacci23}. The thermal analogs of such gels, assembled using colloid-polymer mixtures, have received wide attention \cite{Poon00, Weitz08, Zaccarelli07, Schall20, Tanaka19}, and the emergence of rigidity in these systems is well studied. The role of percolating networks, particularly the directed percolation and isotropic percolation scenarios \cite{Schmiedeberg16, Schall20, Tanaka20,Tanaka19}, is well appreciated. However, there is limited discussion about percolation of colloidal assemblies in active liquids where the colloids form dynamic clusters \cite{Mazza22, Nandi22, Berthier14}. 

Here, we present the findings of an experimental investigation of the percolation transition in a system of non-sticky colloids suspended in active liquids of chiral E. coli. The experiments are conducted in close proximity to a solid-liquid interface, where the swimmers exhibit chiral trajectories. Two notable differences between our experiments and those presented in \cite{Palacci23} are that our spheres are athermal and non-sticky. In our system, the effective interaction between the colloidal particles arises from the activity of the bath, leading to the formation of dynamic clusters \cite{Fakhri22, Chikkadi23, Mishra18}, in contrast to the irreversible aggregates formed in \cite{Palacci23}. Furthermore, the interactions of swimmers with the colloidal particles break the chiral symmetry of colloids, resulting in clusters with persistent rotations. It hinders the orientational ordering of colloids within the clusters, leading to distinct structural features compared to equilibrium colloid-polymer mixtures. The clusters in our system grow with increasing density of colloids and eventually span the system size. This phenomenon displays several hallmark features of equilibrium percolation \cite{Stauffer18, Stanley81, Grest85}. However, the presence of activity and chirality gives rise to exponents that deviate from standard percolation models.  

\section{Experimental set-up}
In our study, we have realized an active-passive mixture by dispersing polystyrene beads of size 15$\mu m$ in active suspensions of E. coli. The procedure for preparing the active fluid is same as the one detailed in our prior publication \cite{Chikkadi23}. Throughout our experiments, we maintained a constant bacterial density equivalent to $10b_0$, where $b_0 = 6 \times 10^9$ cells/ml, unless specified explicitly. To investigate the percolation transition, we systematically varied the colloidal area-fraction from $\phi= 0.1-0.6$. The colloidal beads used in our study are sufficiently large, which makes the Brownian fluctuations negligible. The dynamic behavior of colloidal beads within the active fluid is exclusively driven by the inherent activity of the fluid medium and their hydrodynamic interactions\cite{Yang20, Peng23}. To facilitate our measurements, we constructed an observation chamber by affixing a circular cavity of 1 cm in diameter and 100 $\mu m$ in depth to a coverslip coated with polyethylene glycol (PEG), using double-sided tape. 

For experimentation, we thoroughly mixed the active fluid and colloidal particles before introducing them into the sample chamber, which was subsequently mounted on the microscope stage. Following the loading procedure, we allowed the system to sediment, evolve and, equilibrate for a duration of 1.5 hours before commencing our measurements. All the measurements were made in the stationary state, we captured snapshots of the sample at a speed of 10 FPS over a total duration of 500 seconds.

\section{Results}

\subsection{Chiral active liquids and chiral colloidal assemblies}
\begin{figure}[ht]
\begin{tabular}{cc}
\includegraphics[width=0.22\textwidth]{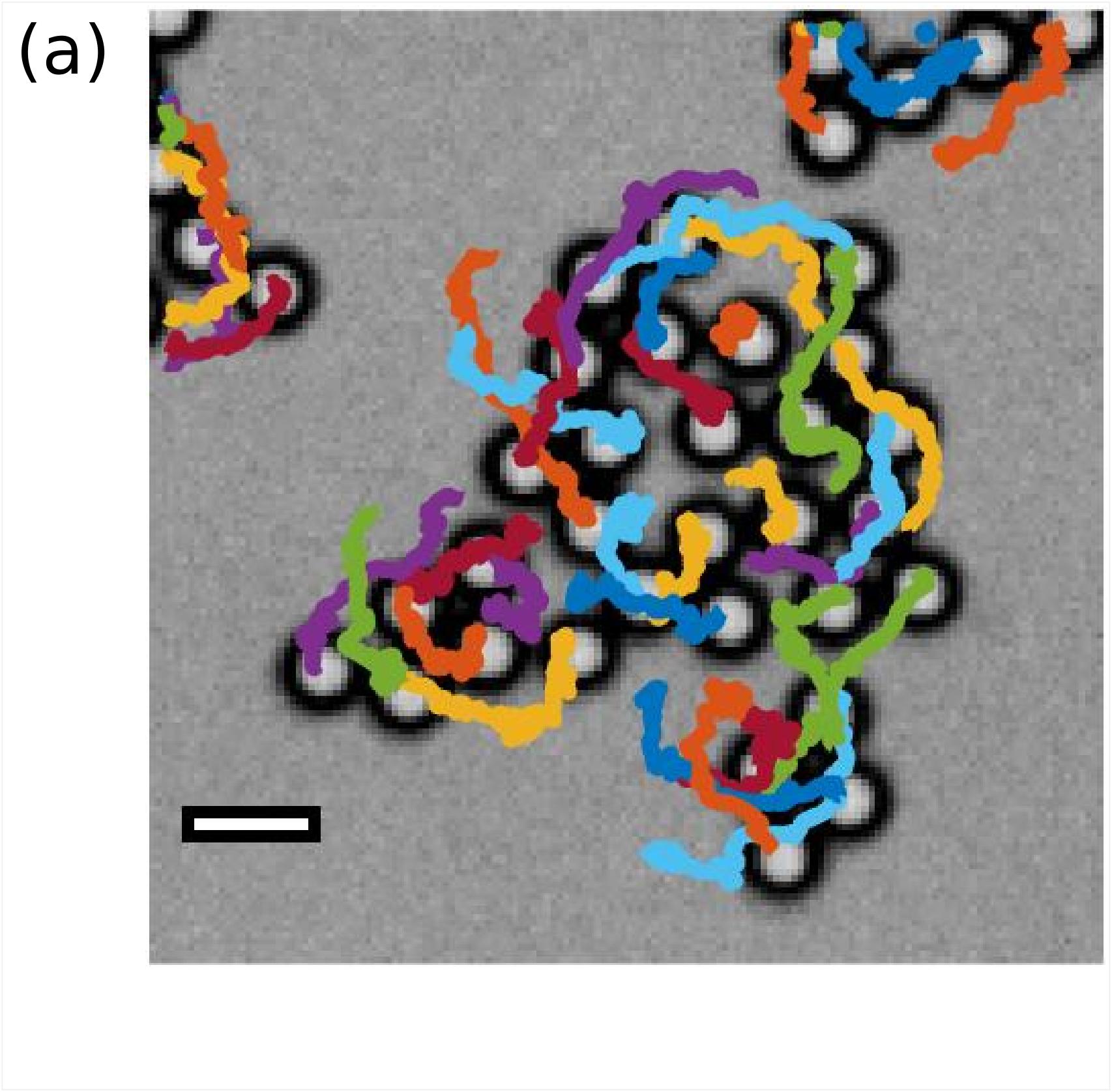}&
\includegraphics[width=0.22\textwidth]{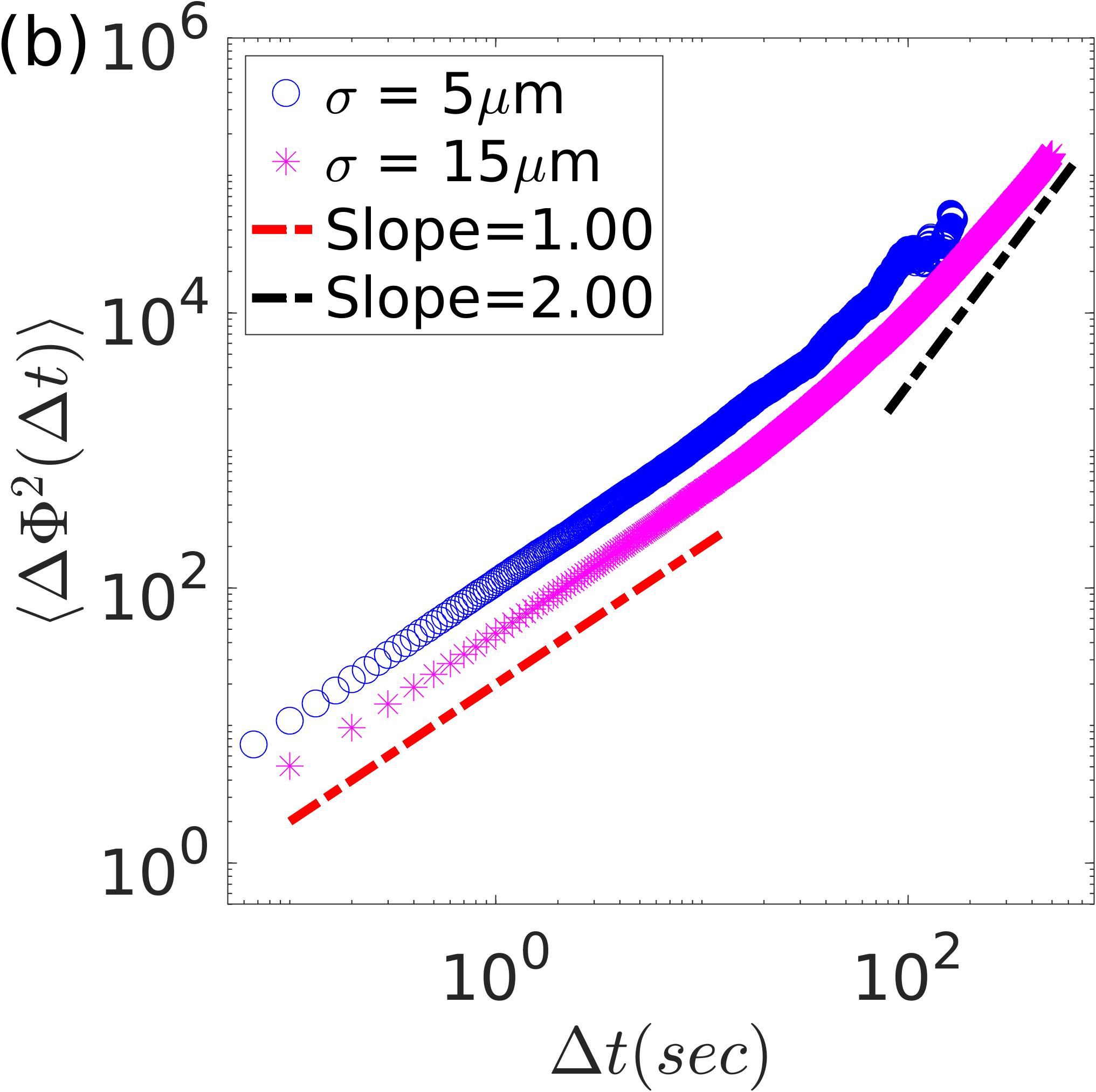}
\end{tabular}
\caption{Broken chiral symmetry of colloidal clusters. (a) Persistent rotation of colloidal clusters are shown by overlapping the particles and their trajectories using different colors. The scale bar measures 20$\mu m$.(b) Mean square angular displacement $\langle \Delta\Phi^2(\Delta t) \rangle$ for $5\mu m$, and $15\mu m$ particle size, red and black dashed lines with slope = 1 and 2, respectively, are shown for reference.}
\label{fig1}
\end{figure}

The experiments detailed in this article are conducted in close proximity to the lower boundary of the chamber, specifically at the solid-liquid interface. It is well-established that E. coli break their chiral symmetry when they swim near a boundary, resulting in curved trajectories displaying clockwise rotation  \cite{Lauga14, Stone06}. The colloidal particles employed in our investigation are $15 \mu m$ in size and are about 5 times larger compared to the size of bacteria, so they experience a noticeable net torque due to interactions with bacteria, thereby breaking their chiral symmetry as well \cite{Leonardo10, Aranson10, Palacci23}. The rotation of individual colloidal particle can be visualised by suspending a Janus particle as shown in the supplementary video SV1. Furthermore, the effective interactions between colloidal particles in active liquids leads to formation of dynamic clusters \cite{Fakhri22, Chikkadi23}, which show persistent rotations, see supplementary video SV2 and SV3. To illustrate the rotational motion, we have plotted the trajectories of a few colloidal paticles in Fig.~1a, where the chiral motion is evident from the curved trajectories. We further compute their mean square angular displacement, which is an angular analogue of the mean square displacement. The unbounded mean square angular displacement is defined as  $\left<\Delta\Phi^2(\Delta t)\right>=\langle [\Phi(t+\Delta t) - \Phi(t)]^2 \rangle$ \cite{Weeks12}, where $\Phi(t)=(1/2\pi)\int_{0}^{t}\Delta\theta(t')dt'$ is the total angular displacement that is integrated over the instantaneous angular displacements $\Delta \theta(t)$. The results are depicted in Fig.~1b for different sizes of colloidal particles $\sigma = 5~\mu m$ and $15~\mu m$ at a bacterial density of $10b_0$. The observed angular motion of particles reveals diffusive characteristics with a slope approaching 1 at short time intervals. However, over larger time scales, the larger particles clearly exhibit a persistent angular rotation, characterized by a slope nearing 2. This effect is prominent for larger ($15 \mu m$) particles, compared to the smaller ones ($5 \mu m$) that show super-diffusive chiral motion on measurement time scales. This chiral motion of colloidal particles and bacteria leads to interesting features as the clusters grow and percolate at a critical area-fraction of colloids.

\subsection{Clusters of colloidal particles}
\begin{figure}[ht]
\centering
\begin{tabular}{ccc}
\includegraphics[width=0.14\textwidth]{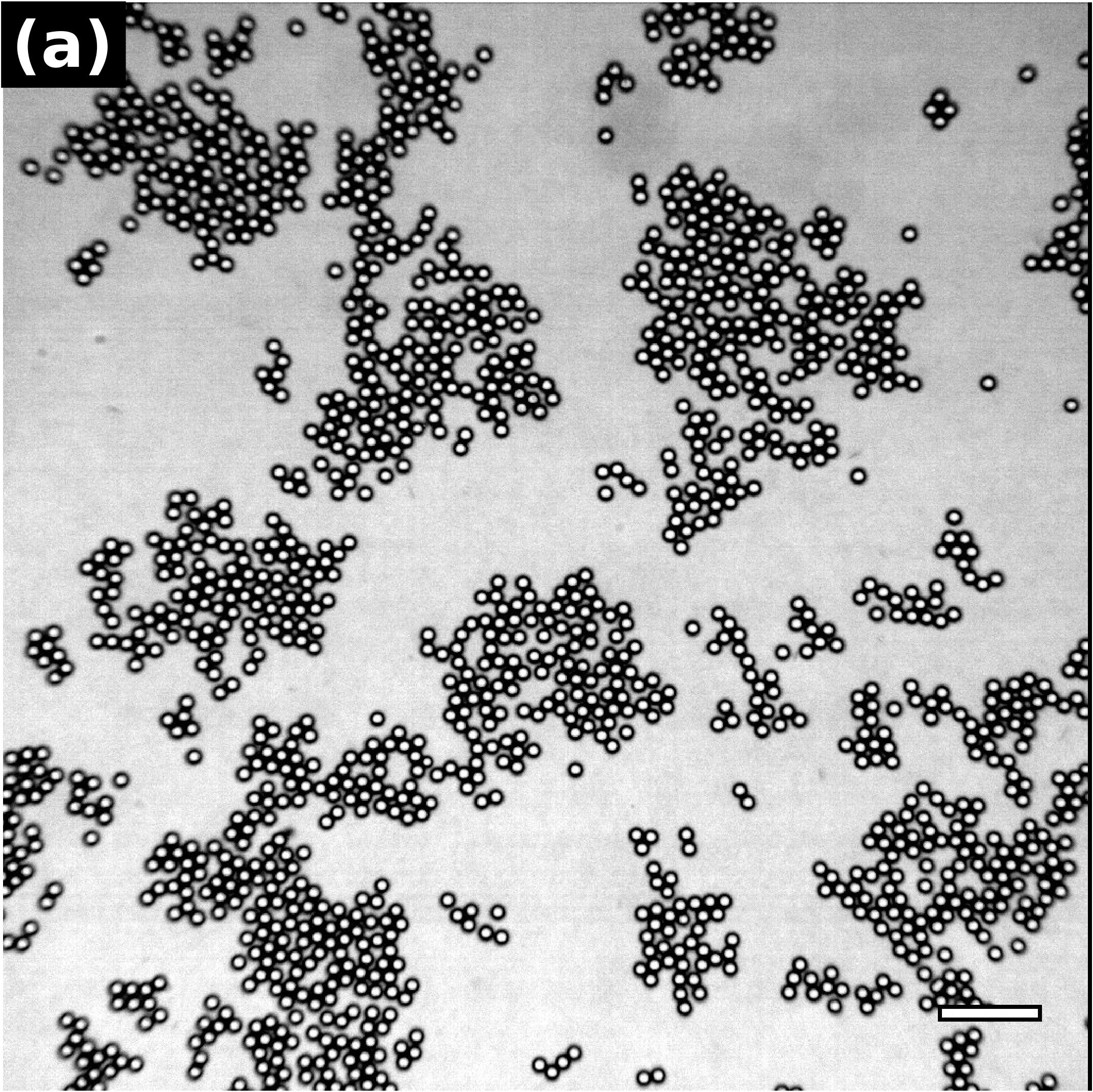}&
\includegraphics[width=0.14\textwidth]{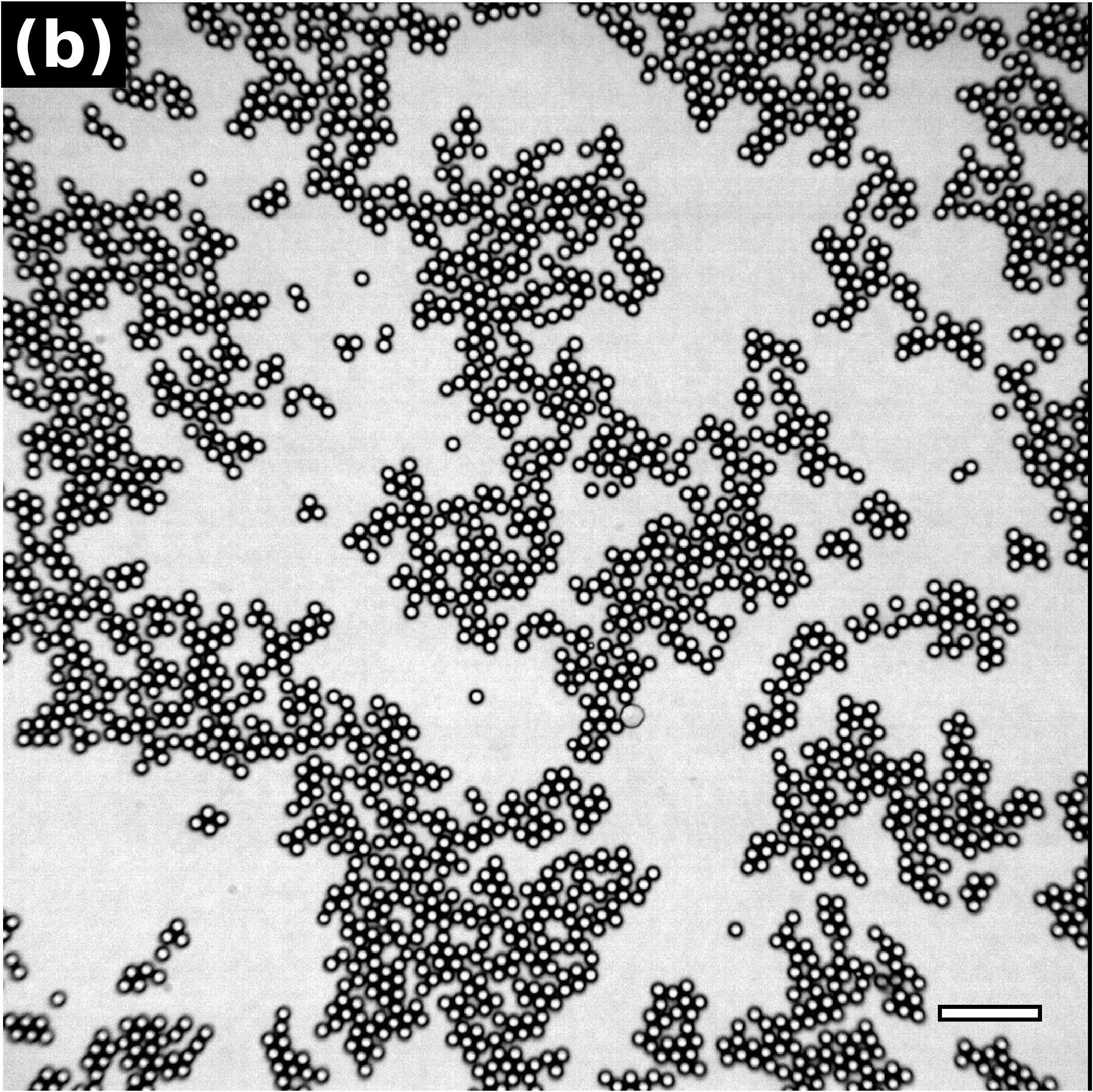}&
\includegraphics[width=0.14\textwidth]{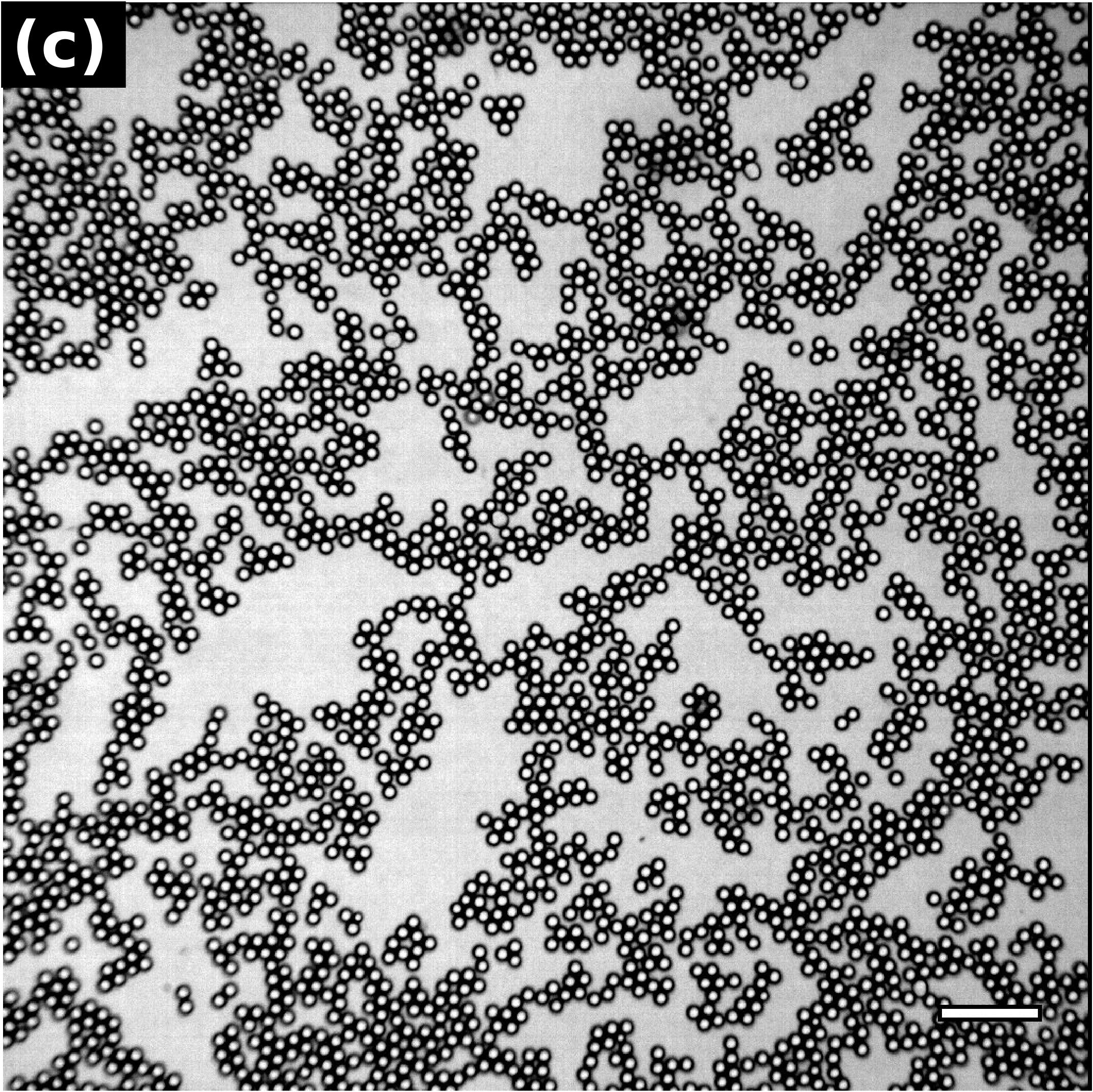}\\
\includegraphics[width=0.14\textwidth]{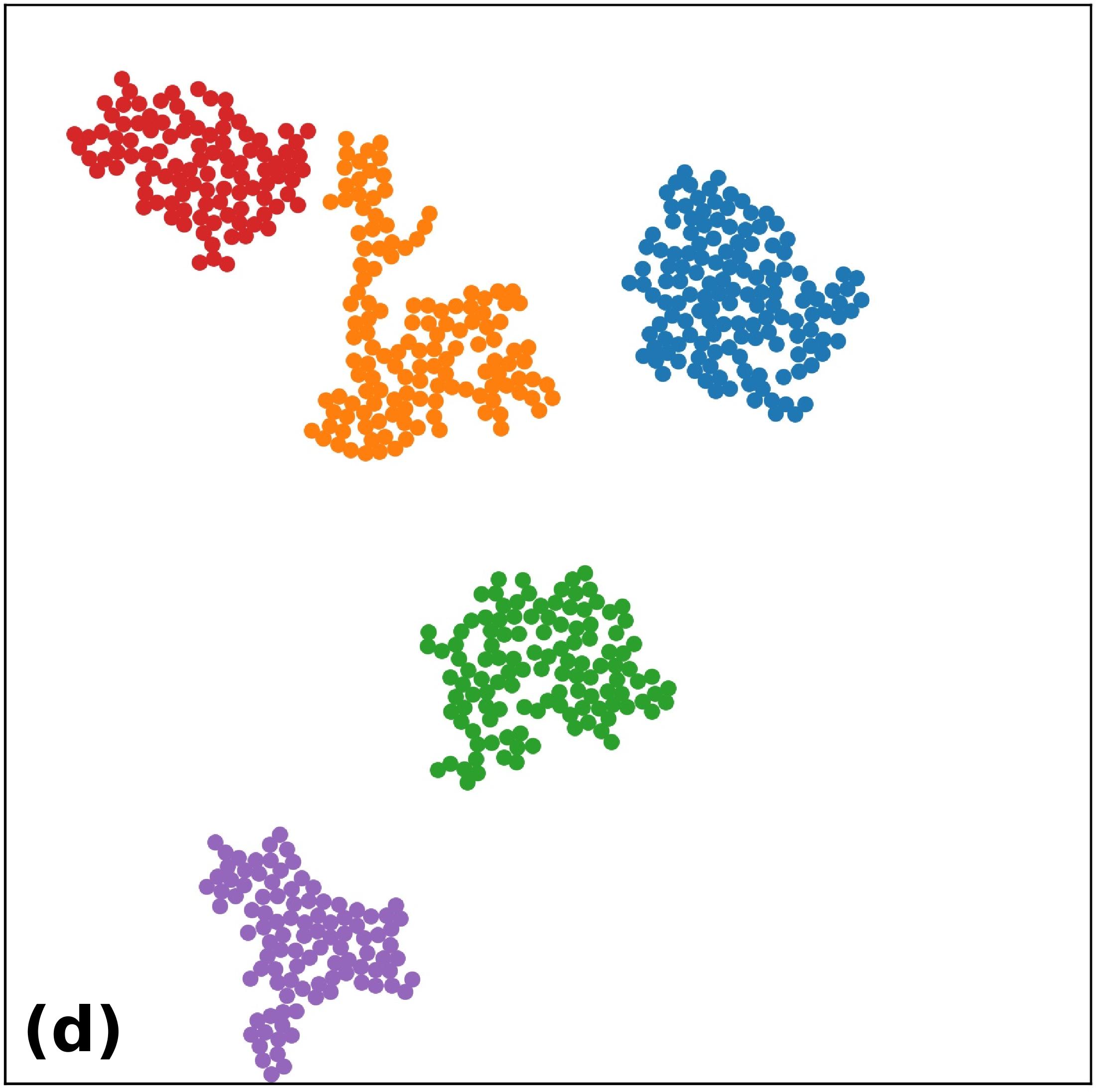}&
\includegraphics[width=0.14\textwidth]{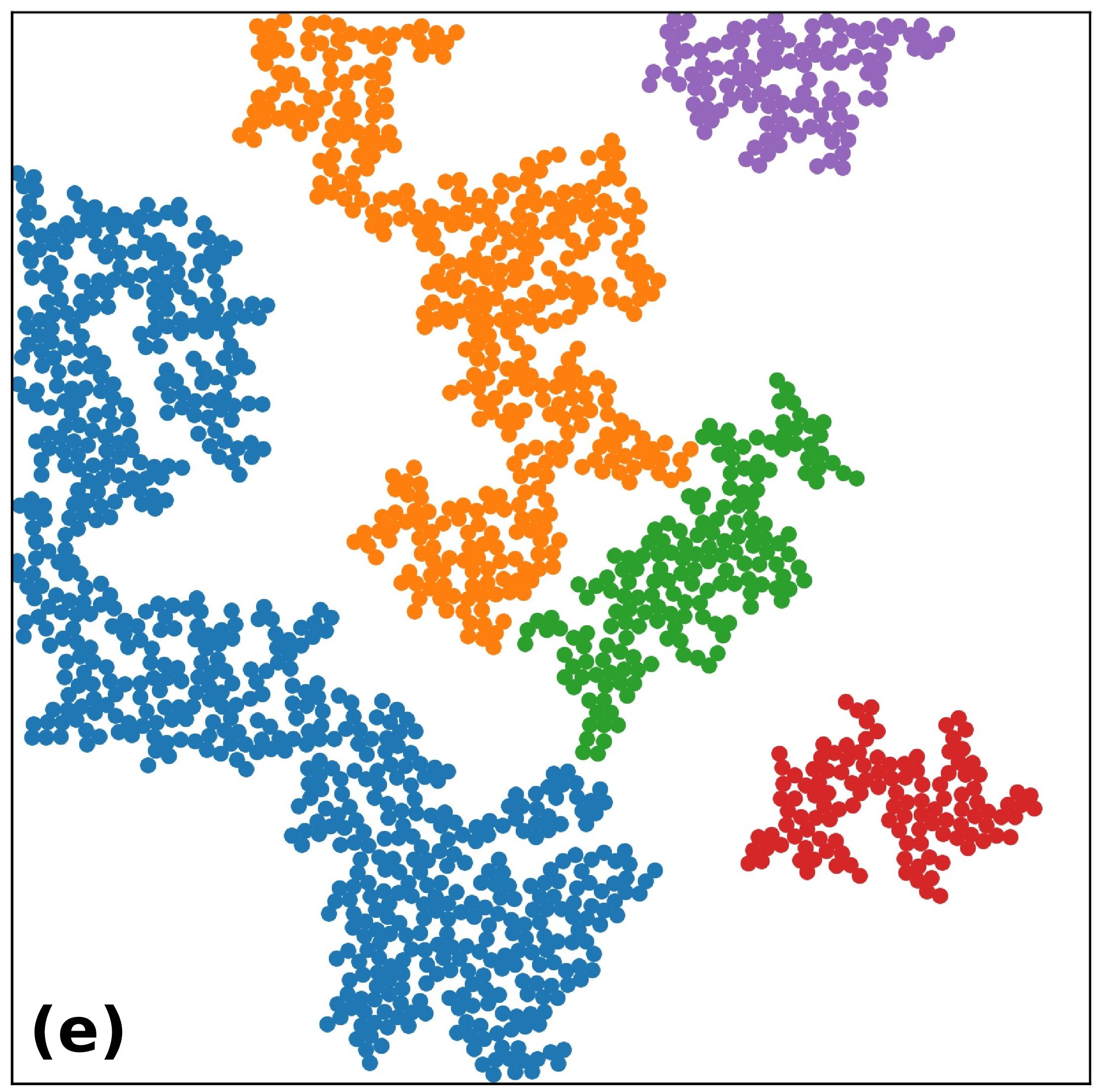}&
\includegraphics[width=0.14\textwidth]{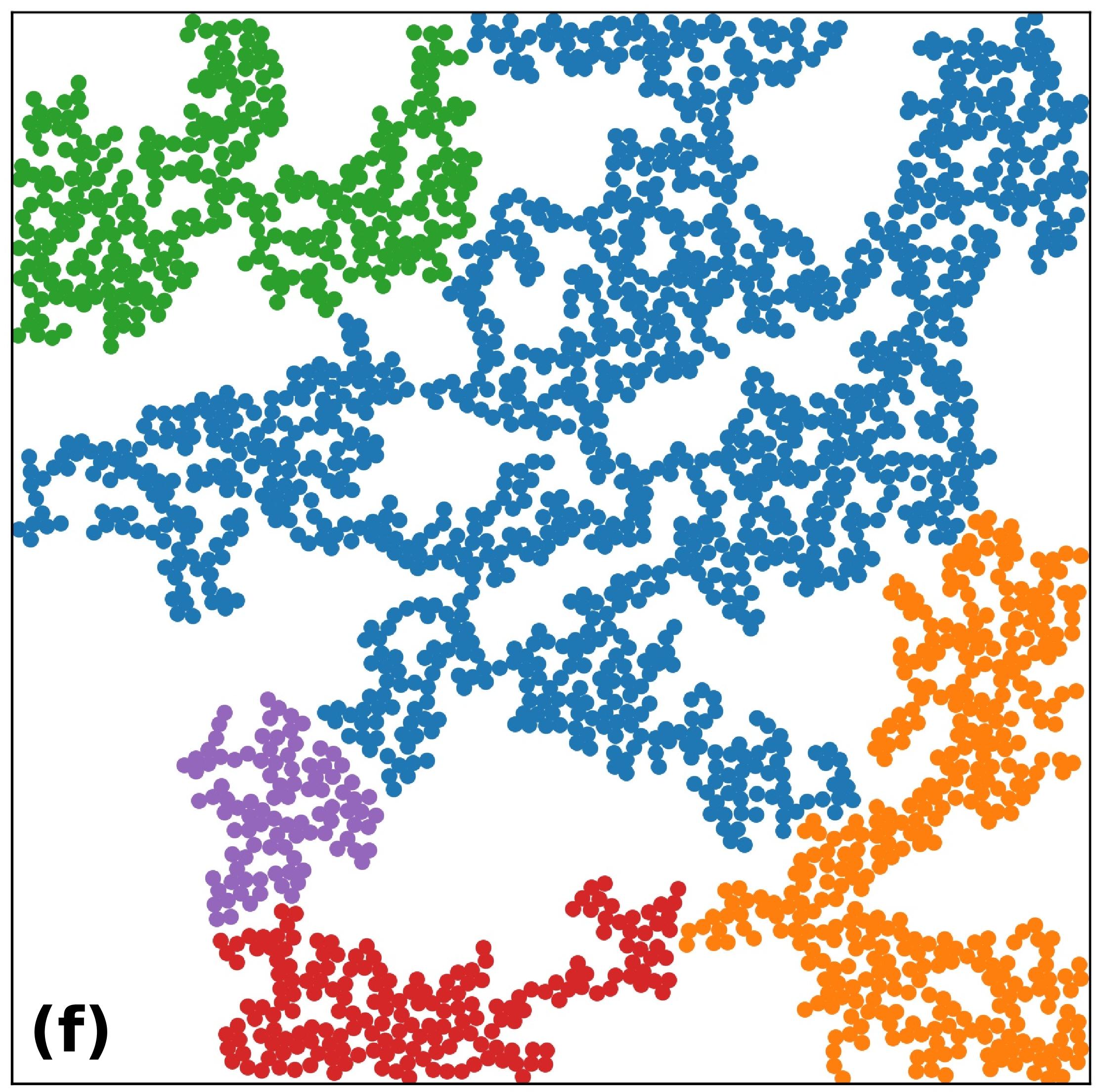}
\end{tabular}
\caption{Percolation of colloidal clusters. Panels (a)-(c) : Bright field images of the colloids in active liquid of E. coli at (a) $\phi$ = 0.2, (b) $\phi$ = 0.3, and (c) $\phi$ = 0.4. The scale bar shown in the images is 100$\mu m$. Panels (d) - (f) : First five largest clusters in the bright-field images shown in the panels (a) - (c). Blue, orange, green, red, and violet are the colors of first to fifth largest cluster, respectively.}
\label{fig2}
\end{figure}

We next display the bright-field images of colloidal assemblies in active liquids in Fig.~2. The panels Figs.2(a)-2(c) correspond to several area-fractions of colloids with $\phi=0.2, 0.3,$ and $0.4$, respectively. The scale bar in these images measures 100 $\mu m$. This is reminiscent of the sol-gel transition in colloidal suspensions with attractive interactions \cite{Zaccarelli07}. The clusters at $\phi=0.2$ in Fig.~2(a) appear to be compact and isolated. However, as $\phi$ increases, the clusters grow in size and become ramified, eventually spanning the entire field of view. To further illuminate these observations, we have depicted the first five largest clusters of Figs.~2(a)-2(c) in Figs.~2(d)-2(f). In these images, the colors blue, orange, green, red, and violet correspond to the first to fifth largest clusters, respectively. The Figs.~2(d)-2(f) provide better insights into the evolution of these clusters with changing $\phi$. The first largest cluster, depicted in blue, gradually grows with increasing $\phi $ and becomes increasingly ramified with a fractal-like structure. Further, the second largest cluster, shown in orange, experiences growth initially but later diminishes in size as most particles are absorbed into the first largest cluster.

\subsection{Spanning probability and percolation threshold}

\begin{figure}[ht]
\begin{tabular}{ll}
\includegraphics[width=0.22\textwidth]{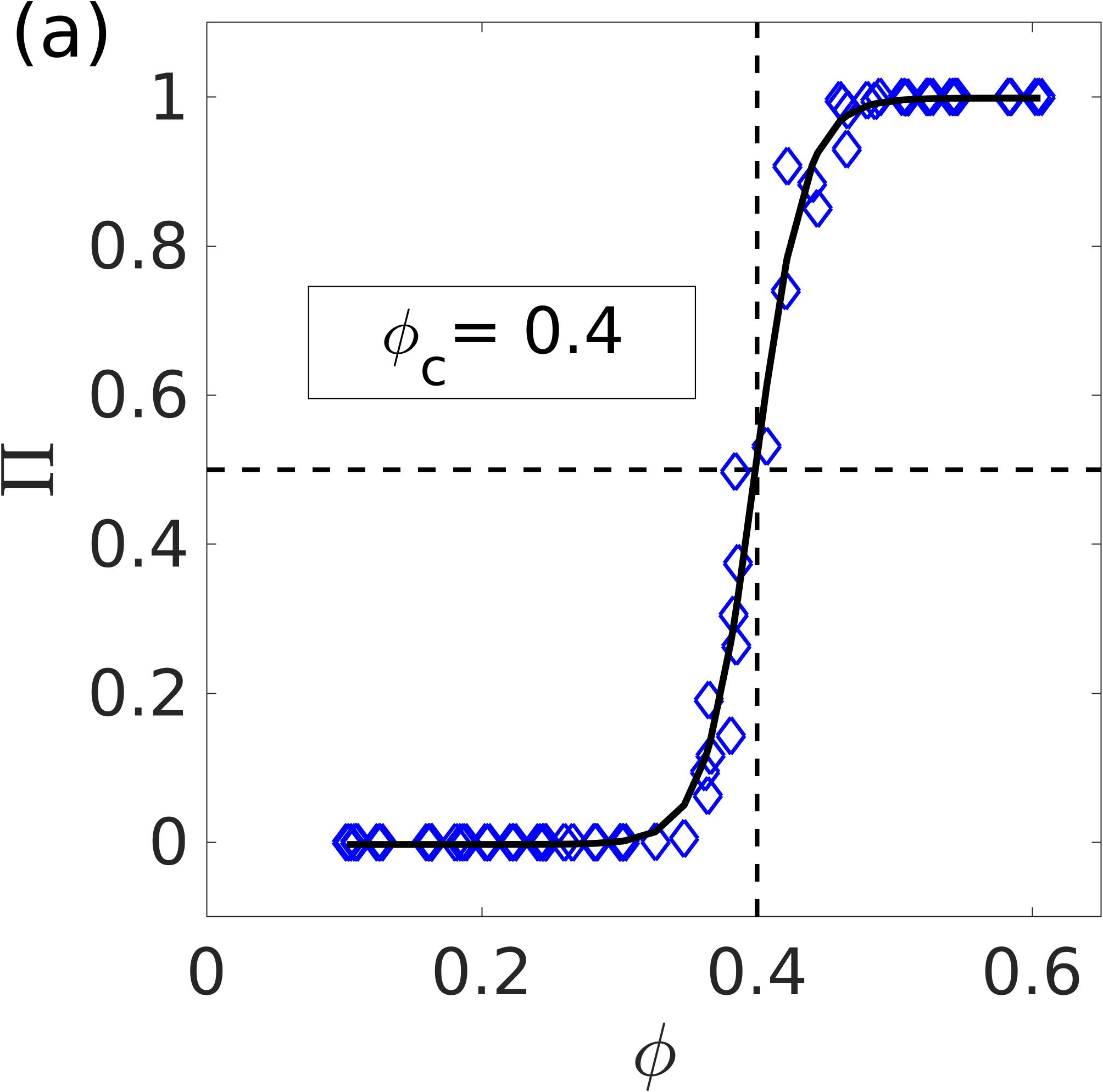}&
\includegraphics[width=0.23\textwidth]{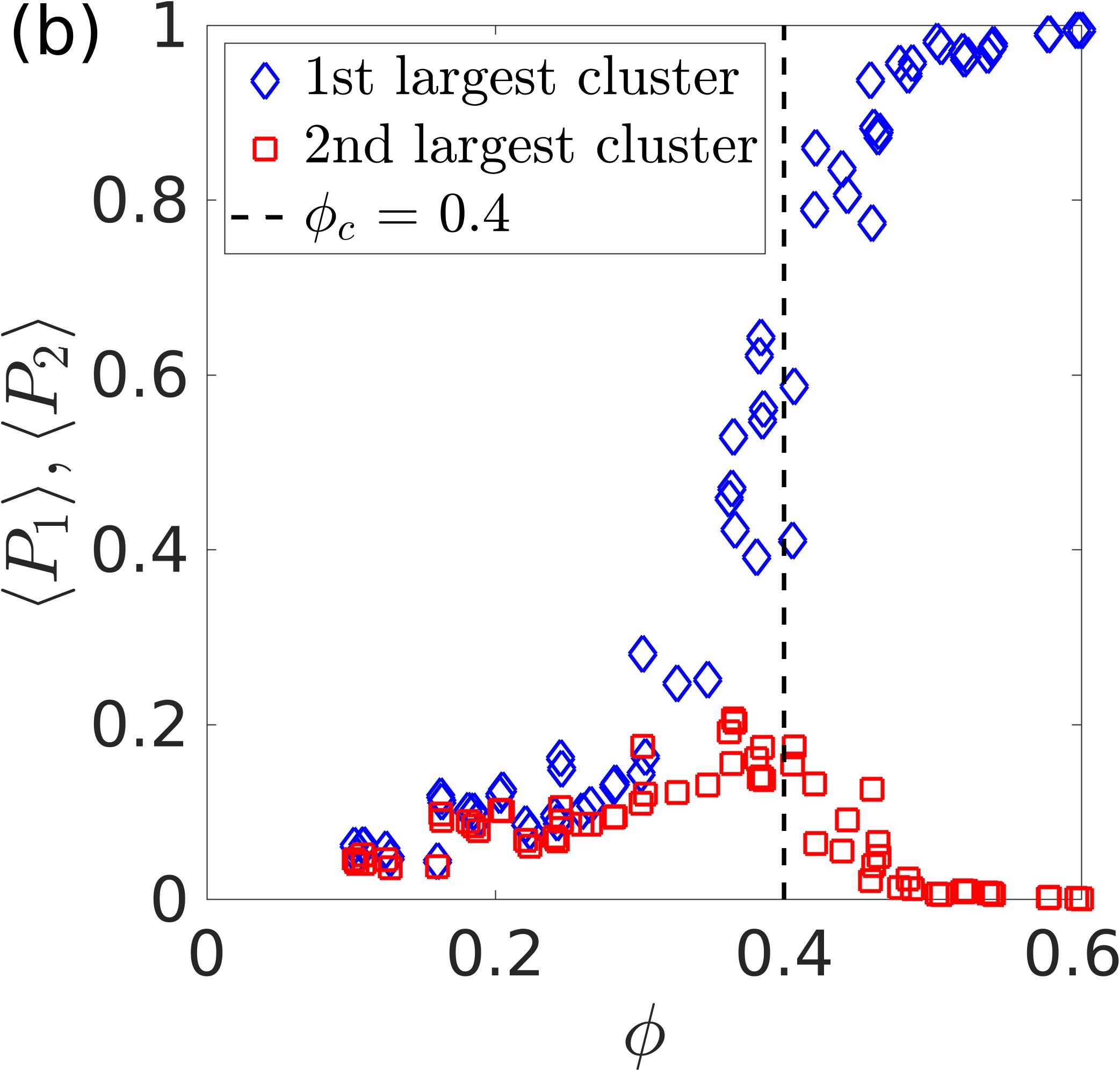}
\end{tabular}
\caption{(a) The spanning probability of the clusters at area-fractions of colloidal particles ranging from $\phi = 0.1-0.6$ and a bacteria concentration of $10b_0$. The black curve is guide to the eye. Horizontal and vertical dashed line are drawn to indicate the percolation threshold $\phi_c$=0.4 at $\Pi = 0.5$. (b) The fraction of particles in the first $\langle P_1 \rangle$ and second $\langle P_2 \rangle$ largest clusters at several area-fraction of colloids. The dashed line shows the percolation threshold $\phi_c = 0.4$.}
\label{fig3}
\end{figure}

As evident from the results in Fig.~2, an increase in the area fraction $\phi$ leads to the growth of the first largest cluster, which eventually spans the entire field of view. In our pursuit of investigating percolation properties, we first determine the percolation threshold, denoted as $\phi_c$. To ascertain this threshold, we employed the concept of spanning probability, denoted as $\Pi$. Spanning probability is the fraction of configurations in which the clusters span in both horizontal and vertical directions for a given $\phi$. This procedure is repeated at several area-fractions $\phi$ of colloidal particles. The critical density or the percolation threshold $\phi_c$ is defined as the area-fraction where the spanning probability $\Pi$ attains a value of 0.5 \cite{Ziff92, Torquato90}. 

Figure 3(a) shows the plot of spanning probability $\Pi$ as a function of $\phi$. The blue diamonds represent the experimental data points, while the solid black curve is guide to the eye. Additionally, the black dashed horizontal and vertical lines gives the intersection point between the spanning probability curve and the $\Pi=0.5$ curve. From this intersection point, we determined the percolation threshold to be $\phi_c \approx 0.4$. Varying the concentration of the bacteria $b_0$ shifts the percolation threshold, it moves to the higher values of $\phi$ with decreasing concentration of swimmers. This effect is shown in Fig.~9 in the appendix section. Moving to Fig.~3(b), we explore the fraction of particles in the first and second largest clusters, denoted as $\langle P_1\rangle$ and $\langle P_2\rangle$, respectively, as a function of $\phi$. The experimental data points for $\langle P_1\rangle$ and $\langle P_2\rangle$ are represented by blue diamonds and red squares, respectively. The black dashed line is drawn at $\phi_c = 0.4$ to denote critical density. In Fig.~3(b), we observe that the fraction of particles in the first largest cluster $\langle P_1\rangle$ increases and eventually saturates as $\phi$ increases. In contrast, the fraction of particles in the second largest cluster $\langle P_2\rangle$ initially increases until $\phi_c$ and decreases beyond. This behavior is a signature of percolation transition, which has been used for defining $\phi_c$ in other studies.\cite{Schall20, Schall17}.

\subsection{Cluster size distribution and radius of gyration}
The percolation transitions are associated with scale-free size distribution of clusters, with no characteristic cluster size at $\phi_c$. To investigate these aspects of cluster formation in our experiments, we investigate their size distribution. We identify clusters by adopting a fixed cutoff distance for connected particles, set at $1.2\sigma$, which is based on the first minima of the pair correlation function g(r). The cluster size distribution denoted as $n_s^*$ is defined as $n_s^*=n_s/\sum n_s$ where $n_s$ is the frequency of clusters of size $s$. 

\begin{figure}[h]
\begin{tabular}{cc}
\includegraphics[width=0.23\textwidth]{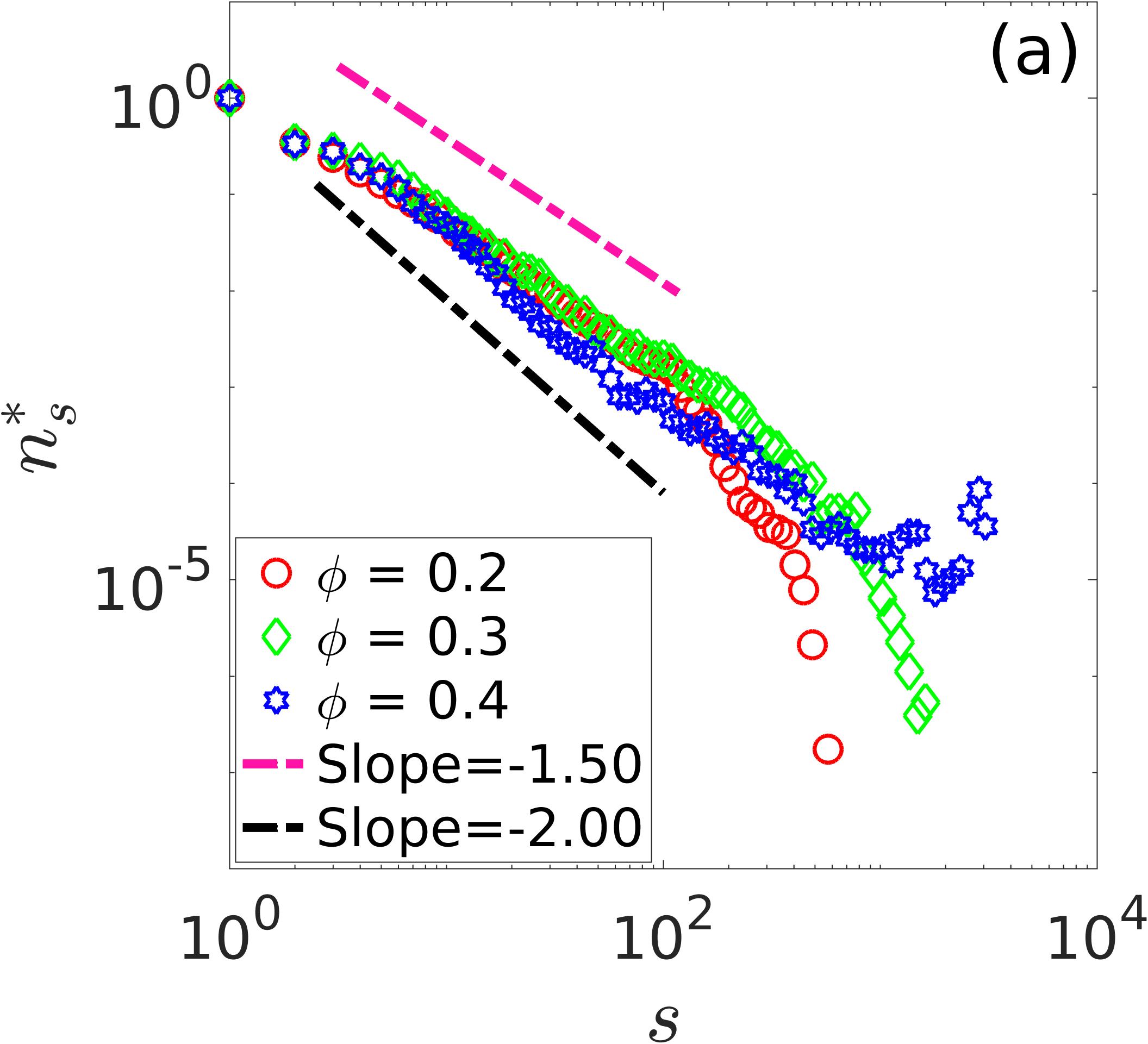}&
\includegraphics[width=0.21\textwidth]{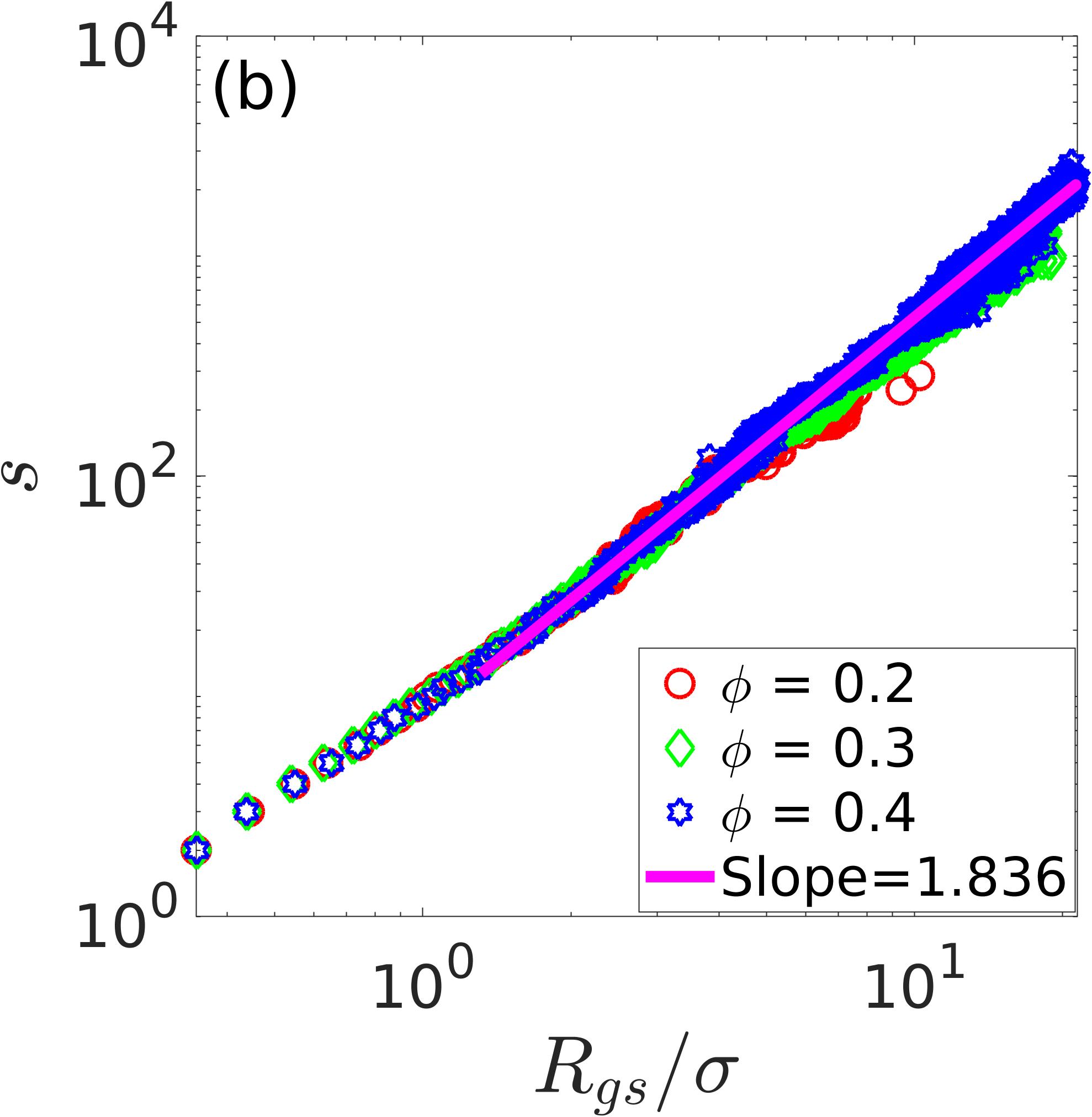}
\end{tabular}
\caption{(a) Cluster size distribution $n_s^*$ as a function of cluster size $s$. Magenta and black dashed line have slope -1.5 and -2, respectively. (b) The scaled radius of gyration $R_{gs}$ of clusters of size $s$. The solid line is a linear fit to the curve corresponding to $\phi_c$ = 0.4, which has a slope of $d_f$ = $1.836\pm 0.040$ for $s>10$.}
\label{fig4}
\end{figure}

The Fig.~4(a) shows the cluster size distribution $n_s^*$ at colloidal densities of $\phi = 0.2, 0.3, $ and $0.4$, denoted by red circles, green diamonds, and blue hexagons, respectively. For the ease of comparison, $n_s^*$ is scaled by $n_1$. The magenta and black dashed lines featuring slopes of -1.5 and -2, respectively, are included for reference. Previous studies \cite{Stauffer18, Christensen02} have shown that cluster size distribution in the vicinity of percolation transition is given by the following expression:
\begin{equation}
n_s^* \propto s^{-\tau } \exp \left( -\frac{s}{s_{\xi}} \right), \text{        for  } s\gg 1,
\end{equation}
where $s_{\xi}$ represents the cutoff length scale of cluster size, and $\tau$ denotes one of the critical exponents. The distribution $n_s^*$ exhibits a power-law dependence with an exponential cutoff below $\phi_c$, however, as $\phi$ approaches $\phi_c$, it approaches a power-law indicating the absence of a characteristic length scale leading to $n_s^*\sim s^{-\tau}$. According to the continuum percolation theory of disks (CPT), the exponent $\tau=-2.0\pm0.1$ in two dimensions \cite{Stanley81} . From Fig.~4(a) it is clear that for $\phi$ below $\phi_c$, the $n_s^*$ has nature of power law decay with exponential cutoff, as $\phi$ approaches to $\phi_c$, it has only power-law dependence. The exponent determined from the best-fit procedure yields a value of $\tau=-1.74\pm 0.12$ at $\phi_c\sim 0.4$, which deviates from the predictions of CPT. Interestingly, the nonequilibrium structure formation in conserved mass models with diffusion, aggregation and fragmentation displays phase transition beyond a critical density $\rho_c$ \cite{Barma98}. The mean-field model predicts $P(s)\sim e^{s/s^*}s^{-3/2}$, where $s$ is the cluster size, below $\rho_c$. However, at $\rho_c$, the cluster size distribution has a power-law form  $P(s)\sim s^{-5/2}$. The inclusion of activity in such models could lead to insightful results.  

We turn our attention to the morphology of the clusters. It appears from Fig.~2 that the clusters acquire a fractal-like structure on approach to percolation transition. This can be tested by calculating the radius of gyration $R_{gs}$ and the associated fractal dimension $d_f$. The radius of gyration, $R_{gs}$, is defined as \cite{Stauffer18}:
\begin{equation}
   {R_{gs}^2 =  \frac{1}{2s^2}\sum_{i j}|\mathbf{r}_i - \mathbf{r}_j|^2 },
\end{equation}
where $\mathbf{r}_i$ and $\mathbf{r}_j$ represent position vectors of particles within a cluster of size $s$ \cite{Stauffer18,Christensen02}. Near the percolation threshold, the cluster size $s$ is connected to the radius of gyration by the relation\cite{Stauffer18}:
\begin{equation}
   s\sim R_{gs}^{d_f},\hspace{0.5cm} (\text{for }\phi =\phi_c, s\gg1),
\end{equation}
where $d_f$ is the fractal dimension. For a two-dimensional system, the $d_f=1.894\pm 0.003$ from the CPT \cite{Heuer93}. The data in Fig.~4(b) shows the relation between the radius of gyration and cluster sizes. The red circles, green diamonds, and blue hexagons represent data points for $\phi = 0.2, 0.3,$ and $0.4$, respectively. The magenta solid line corresponds to the fitting of $R_{gs}$ at $\phi_c=0.4$ for $s>10$, and the slope of the fitted curve gives $d_f = 1.836 \pm~0.040$, which appears to be in close agreement with the continuum percolation model \cite{Stanley81}. 

\begin{figure}[!]
\centering
\begin{tabular}{cc}
\includegraphics[width=0.23\textwidth]{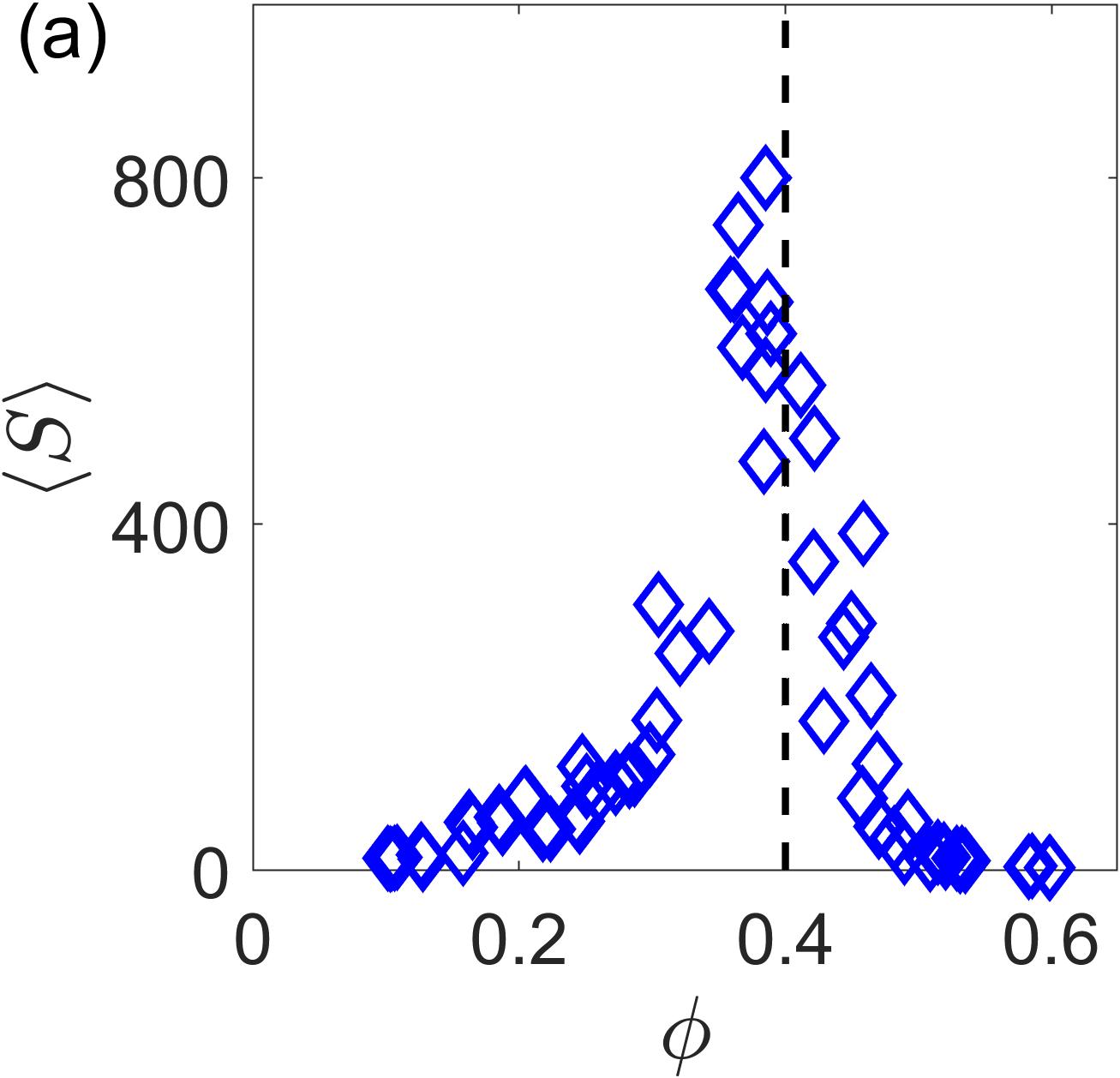}&
\includegraphics[width=0.23\textwidth]{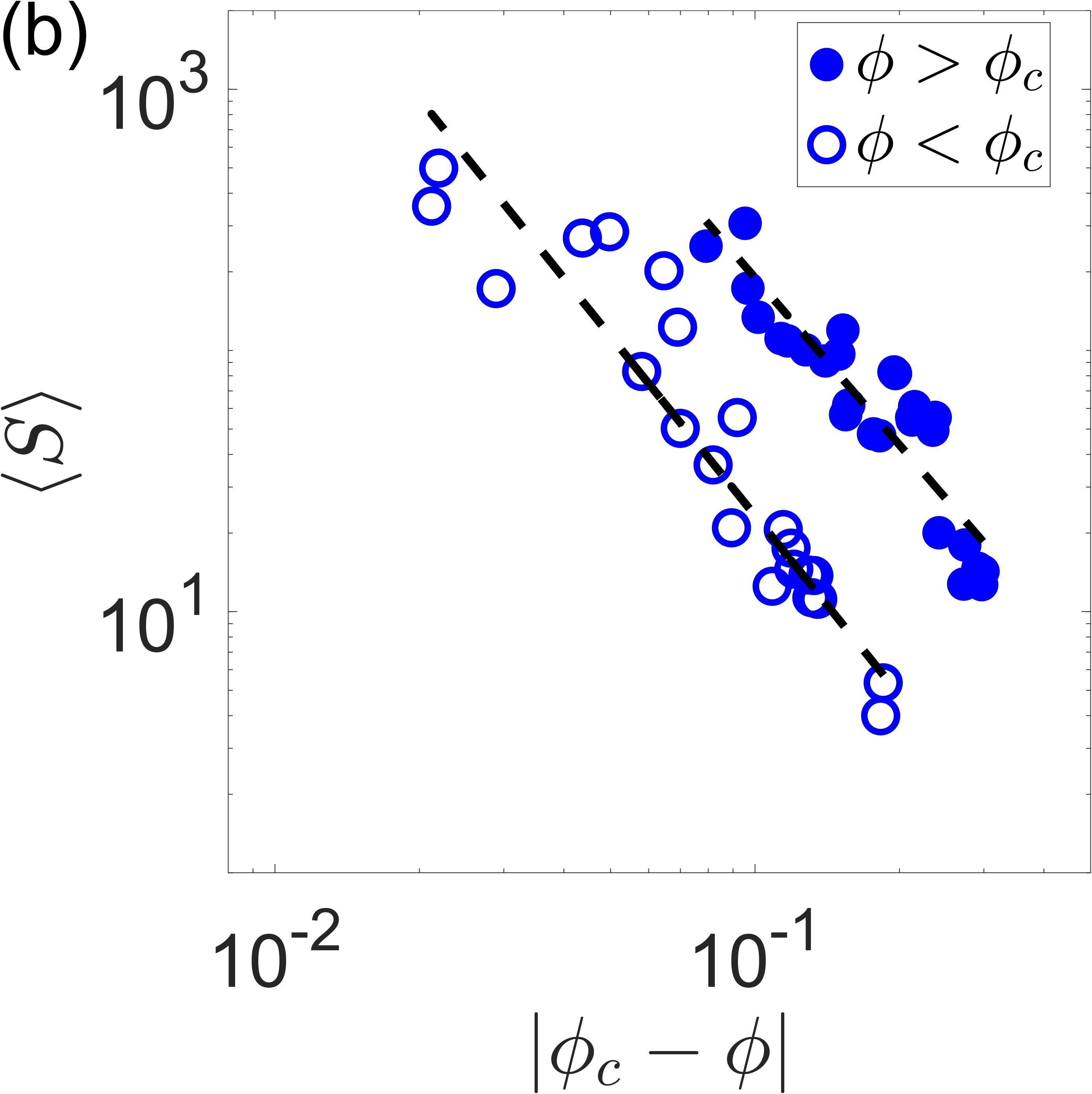}
\end{tabular}
\caption{(a) The mean cluster size $\langle S\rangle$ of colloidal particles as a function of its area-fraction $\phi$. Black dashed line indicate $\phi_c=0.4$. (b) $\langle S\rangle$ as a function of $|\phi_c-\phi|$ for $\phi<\phi_c$ (open circles) and $\phi>\phi_c$ (filled circles). The slope of the best fit line for $\phi<\phi_c$ (open circles) is $\gamma = -2.13\pm 0.16$ and for $\phi>\phi_c$ (filled circles) it is $\gamma = -2.28\pm 0.23$. }
\label{fig5}
\end{figure}
\subsection{Average cluster size $\langle S \rangle$ and correlation length $\langle \xi \rangle$ of clusters}

We proceed further with the analysis of other quantities to establish the nature of percolation transition observed in our experiments. In equilibrium systems quantities like average cluster size $\langle S \rangle$ and correlation length $\langle \xi \rangle$ exhibit power-law scaling with the distance from the critical point. The average cluster size obeys the following relation in the the vicinity of the critical point\cite{Stauffer18}:
\begin{equation}
   {\langle S \rangle \propto |\phi - \phi_c|^\gamma},
   \hspace{2cm}{{\phi \rightarrow \phi_c}} 
\end{equation}
where $\gamma$ is the critical exponent associated with $\langle S \rangle$. The CPT predicts $\gamma=-2.43\pm0.04$ for two-dimensional (2D) systems \cite{Stanley81}. The $\langle S \rangle$ is defined as the second moment of cluster size distribution $n_s^*$ and is give by the following equation\cite{Stauffer18}:
\begin{equation}
   {\langle S \rangle =  \frac{\sum'_{s}s^2n_s}{\sum'_{s}s n_s}}. 
\end{equation}
In the above equation, $s$ represents the size of individual clusters, and $n_s$ denotes the frequency of clusters with size $s$. The notation $\sum'$ signifies that the summation excludes contributions from the spanning cluster or the first largest cluster. The variation of $\langle S \rangle$ with changing area-fraction of colloids across $\phi_c$ is shown in Fig.~5(a). It is important to note that our calculations exclude contributions from spanning clusters, focusing on the behavior of finite-sized clusters. It illustrates a familiar trend close to the critical point. Initially, $\langle S \rangle$ increases as $\phi$ approaches the critical point $\phi_c$ and it decreases beyond $\phi_c$. The Fig.~5(b) shows the variation of $\langle S \rangle$ as a distance from the critical point  $|\phi - \phi_c|$ on a double logarithmic scale. The open circles correspond to $\phi<\phi_c$ and filled circles correspond to $\phi>\phi_c$. The slopes of best-fit lines to the data yields $\gamma = -2.13\pm0.16$ when $\phi<\phi_c$ and the slope is $\gamma = -2.28\pm0.23$ for $\phi>\phi_c$ . In conclusion, our analysis reveals that the critical exponent $\gamma$ associated with $\langle S \rangle$ show deviations from the standard percolation models \cite{Stanley81}. 

\begin{figure}[!]
\centering
\begin{tabular}{cc}
\includegraphics[width=0.22\textwidth]{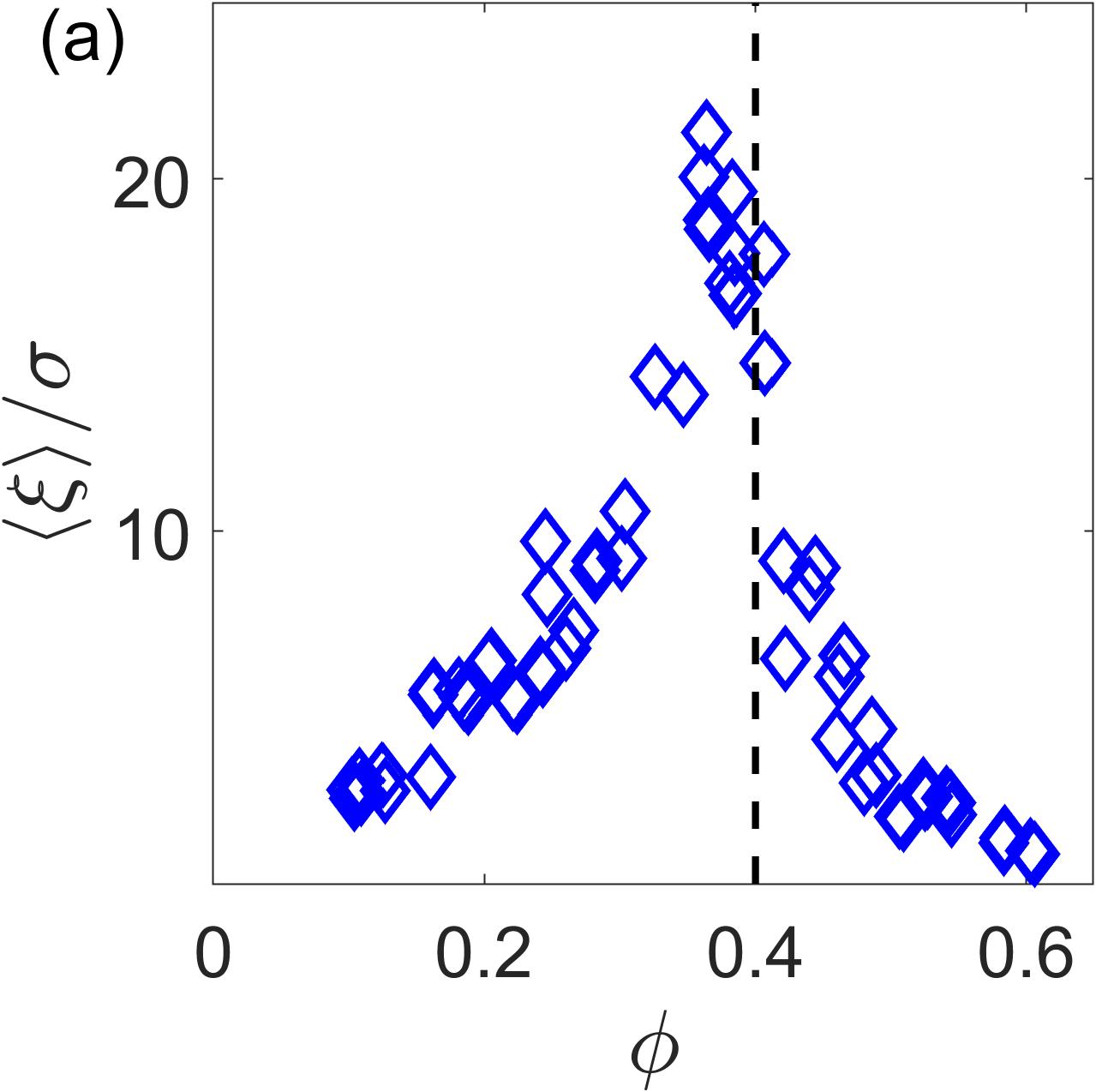}&
\includegraphics[width=0.22\textwidth]{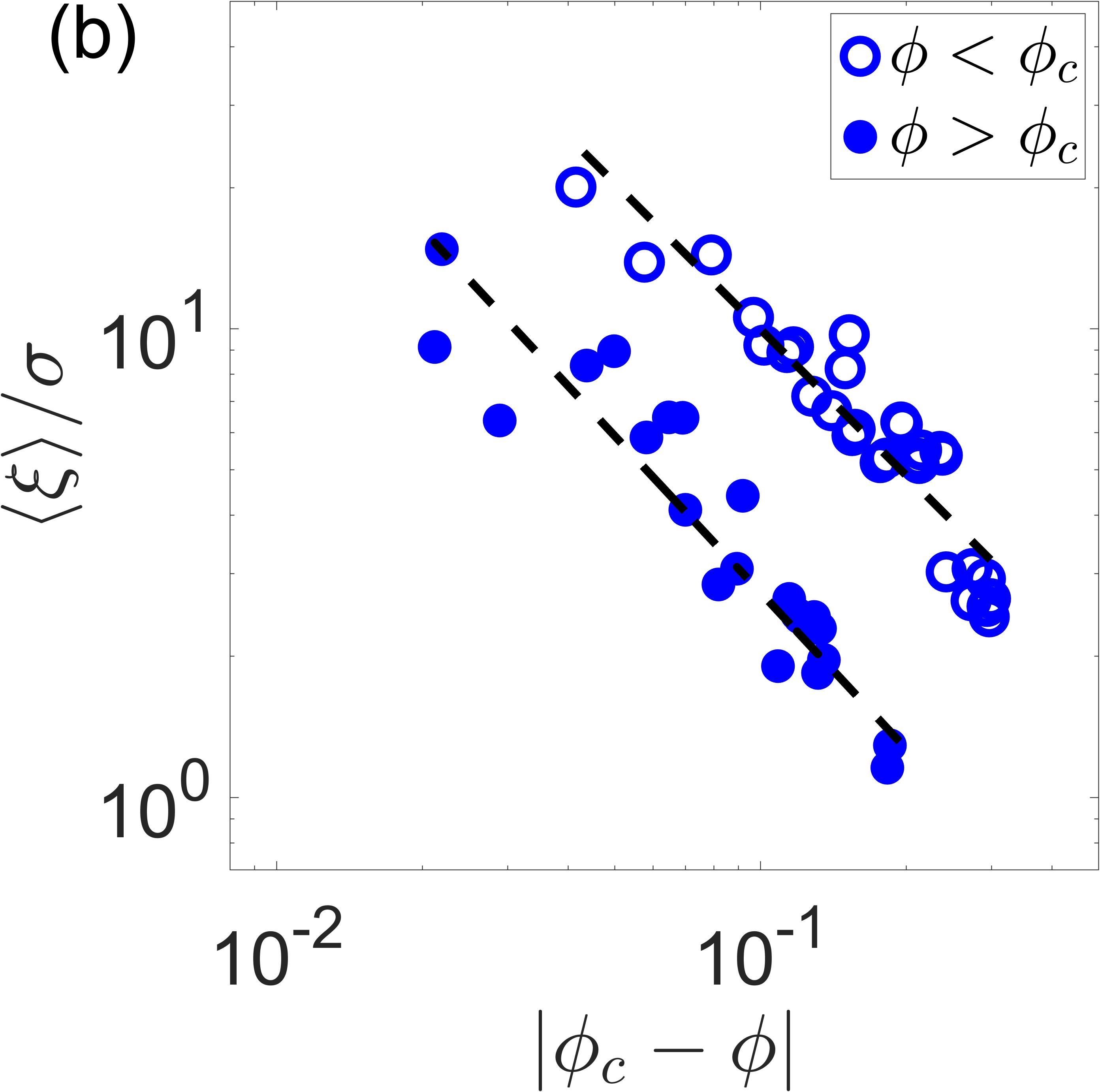}
\end{tabular}
\caption{(a) Correlation length $\langle \xi \rangle$ of the clusters of colloidal particles as a function of its area-fraction $\phi$. Black dashed line denotes the percolation threshold $\phi_c =0.4$. (b) The plot of $\langle \xi \rangle$ as a function of $|\phi_c-\phi|$ for $\phi<\phi_c$ (open circles) and $\phi>\phi_c$ (filled circles). The slope of the best fit line for $\phi<\phi_c$ (open circles) is $\nu = -1.035\pm0.087$ and for $\phi>\phi_c$ (filled circles) it is $\nu = -1.108\pm0.123$.}
\label{fig6}
\end{figure}

Along with $\langle S \rangle$, the equilibrium correlation length $\langle \xi \rangle$ of the clusters is known to obey a scaling relation described by the following equation \cite{Stauffer18}:

\begin{equation}
   {\langle \xi \rangle \propto |\phi - \phi_c|^\nu},
   \hspace{2cm}{{\phi \rightarrow \phi_c}} 
\end{equation}
where the critical exponent $\nu=-1.343\pm0.019$ for a two dimensional system \cite{Stanley81}. The correlation length $\langle \xi \rangle$ is calculated using the equation\cite{Stauffer18}:
\begin{equation}
   {\xi^2 =  \frac{2\sum'_{s}R_{gs}^2s^2n_s}{\sum'_{s}s^2n_s}} 
\end{equation}
where $R_{gs}$ represents the radius of gyration of clusters containing s-particles, and $n_s$ denotes the frequency of clusters with size $s$. Similar to previous definitions, the primed sum excludes contributions from the spanning cluster. The analysis of $\langle \xi \rangle$ in our experiments at various densities of colloids in Fig.~6(a) reveals a similar trend as $\langle S \rangle$, in the neighborhood of $\phi_c$. We can gain a better understanding by presenting the data on a double logarithmic plots in Fig.~6(b). The data in Fig.~6(b) shows the variation of $\langle \xi \rangle$ as a function of distance from the critical point $|\phi_c-\phi|$. The open circles correspond to $\phi<\phi_c$ and filled circles correspond to $\phi>\phi_c$. The slopes of best-fit lines to the data yields $\nu =  -1.035\pm0.087$ when $\phi<\phi_c$ and the slope is $\nu = -1.108\pm0.123$ for $\phi>\phi_c$. The exponents $\nu$ on both the sides of $\phi_c$ is far from the standard continuum model.

The results from our analysis of cluster sizes distribution $n_s^*$, average cluster size $\langle S \rangle$, and correlation length $\langle \xi \rangle$ reveal that their respective exponents $\tau$, $\gamma$, and $\nu$ show deviations from the standard exponents predicted by continuum percolation theory for 2D equilibrium systems. The exponents reported here could have finite size effects and in addition it is difficult to accurately pinpoint the critical density in experiments. Our results motivate novel theoretical models and simulations to determine the true values of exponents. Our current results appear to support recent numerical studies on percolation transition in purely active particle systems \cite{Mazza22, Berthier14} that have reported critical exponents deviating from standard percolation models, hinting at a new universality class of percolation model for active systems. A confirmation of these observation needs further investigation into these aspects.

\subsection{Effect of chiral swimmers on the structure of clusters}

\begin{figure}[h!]
\centering
\begin{tabular}{c}
\includegraphics[width=0.45\textwidth]{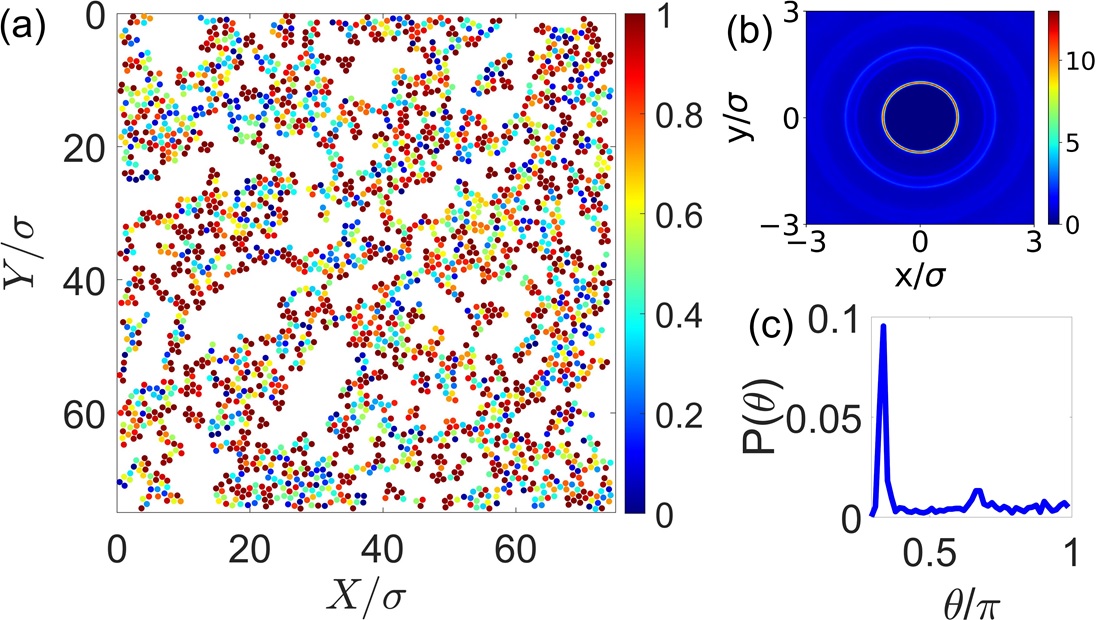}
\end{tabular}
\caption{Structure of the clusters at the colloidal area-fraction of $\phi_c=0.4$. (a) Local bond-orientational-order parameter $\psi_6$ of individual particles is color-coded based on its magnitudes. (b) The two dimensional pair-correlation function g(x,y) of colloidal assemblies. The color indicates the magnitude of g(x,y). (c)The distribution of angle $P(\theta)$ between two successive bonds. }
\label{fig7}
\end{figure}

In this section, we discuss the influence of chirality on the organization of colloids within the clusters. To elucidate this effect, we analyze the microscopic structure of clusters using conventional structural measures such as the local bond orientational order parameter and the pair correlation function. The local bond orientational order parameter is defined as $\psi_6 = \frac{1}{N_j} \sum_{j=1}^{N_j} \text{exp}(6i\theta_{ij})$, where $\theta_{ij}$ is the orientation of the link connecting the two neighboring particles $(i,j)$, and the sum runs over $N_j$ nearest neighboring particles within a cutoff distance of $1.2\sigma$. Figure 7(a) presents the local $\psi_6$ at the percolation threshold $\phi_c$, where the color coding is based on the $\psi_6$ values of the particles. This plot reveals weak orientational ordering of particles in the clusters, further confirmed by the polar plot of the pair correlation function in Fig. 7(b). The plot of $g(x,y)$ demonstrates a liquid-like isotropic structure of the system. We next display the distribution of angles $P(\theta)$ between two successive bonds \cite{Schmiedeberg16} in Fig. 7(c). The plot shows $P(\theta)$ with a prominent peak at $\theta=\frac{\pi}{3}$ and a small one at $\theta=\frac{2\pi}{3}$. These results are robust and persist over a range of densities $\phi$ and bacteria concentration. They are illuminating when interpreted in light of our understanding of equilibrium colloid-polymer mixtures \cite{Poon00, Lekkerkerker11}. Figs. 8(a)-8(c) in the appendix depict the ordering of colloids within clusters of colloid-polymer mixture. The presence of prominent peaks in these plots points to strong directional ordering, which is reported to be responsible for directed percolation transition in colloid-polymer mixtures \cite{Schmiedeberg16, Tanaka20, Tanaka22}. Apparently, the weak structural ordering in colloid-bacteria mixture is a result of activity and chirality of the swimmers, which also alter the nature of the percolation transition.

In summary, we have studied the non-equilibrium percolation of colloidal assemblies in active liquids of chiral swimmers. The colloids form dynamic clusters due to effective interactions mediated by the active bath. The chirality of swimmers strongly influences the dynamics of individual particles and their aggregates, and it leads to persistent clockwise rotations of individual particles and their clusters. It also hinders the local orientational ordering of colloids, which is observed in equilibrium colloid-polymer mixtures. With the increasing density of colloids, the size of the clusters grows and eventually spans the entire system at a critical density $\phi_c=0.4$, which was determined from the spanning probability. Further analysis of various parameters, including cluster size distribution $n_s^*$, radius of gyration $R_{gs}$, average cluster size $\langle S \rangle$, and correlation length $\langle \xi \rangle$, along with their associated exponents $\tau$, $d_f$, $\gamma$, and $\nu$, suggests that colloid-bacteria mixtures displays several hallmark features of percolation transition. However, the exponents show marked deviations from the exponents of standard continuum percolation model. These results need confirmation from finite system size analysis, which is possible with particle simulation methods \cite{Mishra18, Nandi22, Mazza22}. Our experiments reveal richer and tunable structures of colloidal assemblies in non-equilibrium baths, and it should inspire further investigations to unveil novel rheological properties of colloidal assemblies in active liquids.

\section{Appendix}

\subsection{Structure of clusters in colloid-polymer mixture}
In this section, we investigate the structure of clusters in a colloid-polymer mixture, which was prepared utilizing silica beads of diameter of 3.34 $\mu m$ and sodium polystyrene sulfonate (NaPSS) polymer of molecular weight $10^6$ Da. The preparation involved two distinct polymer concentrations, namely $ c_p = 0.1\%$ and $ 0.01\%$, both suspended in deionized (DI) water. The particles sediment to the bottom to form a monolayer of particles, which is evidenced from their gravitational length $l$. For our silica beads, it turns out to be $l\sim 0.02\mu m$, which is much smaller than the size of the particle \cite{Hunter01}.

\begin{figure}[h]
\includegraphics[width=0.45\textwidth]{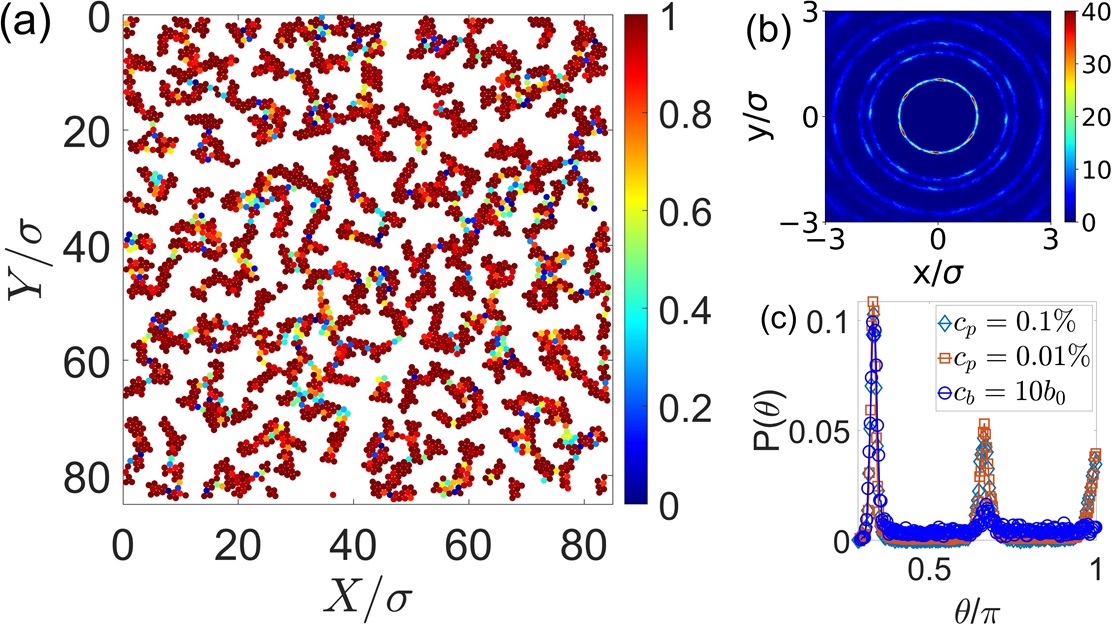}
\caption{Structure of clusters in colloid-polymer mixture at a colloid area-fraction of $\phi=0.40$. (a) The particles are color coded based on the magnitude of their local bond-orientational-order parameter ($\psi_6$). (b) Polar representation of pair-correlation function g(x,y) of colloid-polymer mixture. The color represents the magnitude of g(x,y). (c)The distribution of angle $P(\theta)$ for two different polymer concentrations, $c_p = 0.1\%$ and $0.01\%$, for colloid-polymer mixture, and $c_b=10b_0$ for colloid-bacteria system at $\phi_c=0.4$. }
\end{figure}

The Fig.~8(a) depicts the local bond orientational order parameter ($\psi_6$) for a polymer concentration of $c_p = 0.10\%$ and a colloidal area-fraction of $\phi=0.40$, where the colors illustrates the magnitude of $\psi_6$. Additionally, the Fig.~8(b) presents the two-dimensional pair correlation with the color indicating the magnitude of g(x,y) at same $c_p$ and $\phi$. The local ordering of colloidal particles is evident from the results in these two figures. There is a strong hexatic ordering in Fig.~8(a), while Fig.~8(b) shows distinct peaks in the first two inner rings corresponding to hexagonal symmetry.

The two-dimensional pair correlation g(x,y) is defined as \cite{Swinney10}
\begin{equation}
\tag{S1}
g(x, y) = \frac{1}{\rho} \left\langle\sum_{j \neq i} \delta\left[x\hat{x}_i + y\hat{y}_i - (\mathbf{r}_i - \mathbf{r}_j)\right]\right\rangle_i,
\end{equation}
where \(\delta\) is a Dirac delta function, \(\rho = \frac{N_{\text{total}}}{A}\) is the area density, \(\langle \ldots \rangle_i\) represents the average over all reference particles $i$.

Figure 8(c) shows the distribution of angles $P(\theta)$ between two successive bonds\cite{Schmiedeberg16} for two different polymer concentrations, $c_p = 0.1\%$ and $0.01\%$, at $\phi=0.4$, respectively. This analysis also includes the colloid-bacteria mixture at a cell concentration of $c_b=10b_0$ and a colloidal area-fraction of $\phi=0.4$. The $P(\theta)$ curve displays a prominent peak at $\frac{\pi}{3}$, along with enhanced peaks at $\frac{2\pi}{3}$ and $\pi$. Apparently, such directional ordering is absent in the clusters of colloidal-bacteria system.

\subsection{Effect of varying swimmer density on the spanning probability }

\begin{figure}[h!]
\centering
\includegraphics[width=0.3\textwidth]{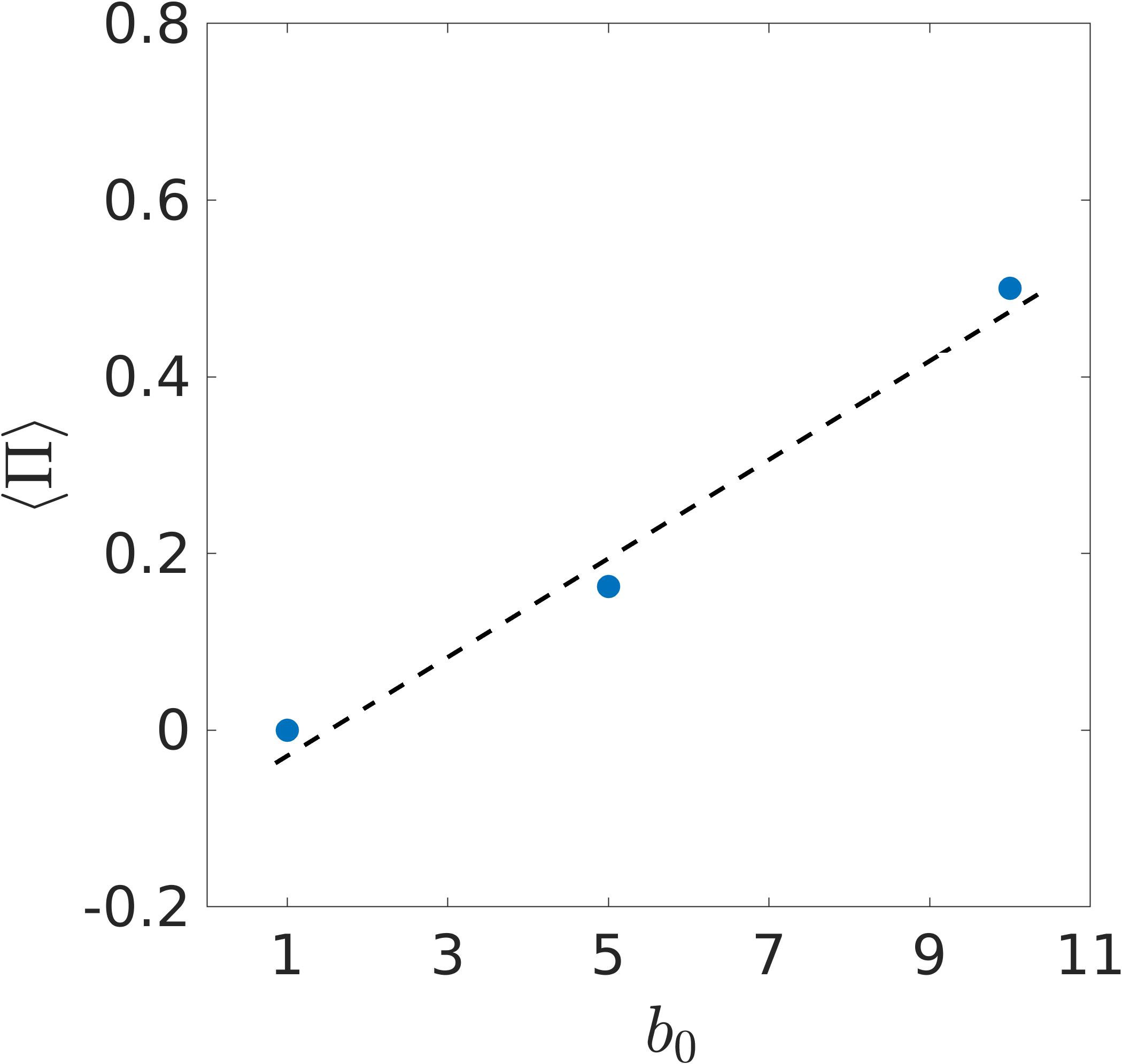}
\caption{Spanning probability of the colloidal clusters at $\phi = 0.4$ and different densities of bacteria $Nb_0$, where $b_0=6 \times 10^9$cells/ml.}
\end{figure}

Figure 9 shows the effect of varying the density of swimmers on the spanning probability of colloidal clusters. The density of bacteria used in all the measurements in the main paper is $10b_0$. Here, we have varied the density of swimmers from $1b_0$ to $10b_0$. The spanning probability increases with increasing concentration of bacteria. This  suggests that the clusters grow in size with increasing density of bacteria \cite{Fakhri22}.


\medskip
%

\end{document}